\documentclass[aps,pra,amsmath,amssymb,nofootinbib,floatfix]{revtex4-2}

\usepackage[T1]{fontenc}
\usepackage[T1]{tipa}
\usepackage{graphicx}
\usepackage{physics}
\usepackage{bm}
\usepackage{times}
\usepackage{natbib}
\usepackage{hyperref}
\usepackage{jabbrv}

\begin{document}

\title{Quantum enhanced distributed phase sensing with a truncated SU(1,1) interferometer}

\author{Seongjin Hong$^{1,2,}$\footnote{shong@cau.ac.kr}, Matthew A. Feldman$^{3,4}$, Claire E. Marvinney$^{3,4}$, Donghwa Lee$^{5}$, Changhyoup Lee$^{6}$, Michael T. Febbraro$^{1}$, Alberto M. Marino$^{3,4,}$\footnote{marinoa@ornl.gov}, and Raphael C. Pooser$^{3,4}$}

\affiliation{$^{1}$Physical Sciences Division, Oak Ridge National Laboratory, Oak Ridge, TN 37831, USA$^{\ddag}$\\
$^{2}$Department of Physics, Chung-Ang University, Seoul 06974, Korea\\
$^{3}$Quantum Information Science Section, Computational Sciences and Engineering Division, Oak Ridge National Laboratory, Oak Ridge, TN 37831, USA$^{\ddag}$\\
$^{4}$Quantum Science Center, Oak Ridge National Laboratory, Oak Ridge, TN 37831, USA$^{\ddag}$\\
$^{5}$Center for Quantum Information, Korea Institute of Science and Technology, Seoul 02792, Korea\\
$^{6}$Korea Research Institute of Standards and Science, Daejeon 34113, Korea}

\thanks{This manuscript has been authored, in part, by UT-Battelle LLC, under
contract DE-AC05-00OR22725 with the U.S. Department of Energy (DOE). The publisher
acknowledges the U.S. government license to provide public access under the DOE Public
Access Plan (http://energy.gov/downloads/doe-public-access-plan).}

\begin{abstract}
In recent years, distributed quantum sensing has gained interest for a range of applications requiring networks of sensors, from global-scale clock synchronization to high energy physics. In particular, a network of entangled sensors can improve not only the sensitivity beyond the shot noise limit, but also enable a Heisenberg scaling with the number of sensors.  Here, using bright entangled twin beams, we theoretically and experimentally demonstrate the detection of a linear combination of two distributed phases beyond the shot noise limit with a truncated SU(1,1) interferometer. We experimentally demonstrate a quantum noise reduction of 1.7 dB and a classical 3 dB signal-to-noise ratio improvement over the separable sensing approach involving two truncated SU(1,1) interferometers. Additionally, we theoretically extend the use of a truncated SU(1,1) interferometer to a multi-phase-distributed sensing scheme that leverages entanglement as a resource to achieve a quantum improvement in the scaling with the number of sensors in the network. Our results pave the way for developing quantum enhanced sensor networks that can achieve an entanglement-enhanced sensitivity.
\end{abstract}

\maketitle

\section{Introduction}
Quantum metrology allows us to estimate an unknown parameter with enhanced sensitivity over classical approaches by exploiting quantum resources \cite{caves1981quantum,giovannetti2004quantum,pirandola2018advances,lawrie2019quantum,RevModPhys.90.035005,RevModPhys.89.035002}. Quantum sensors that use entanglement or other quantum correlations provide a promising platform that has been used in a variety of sensing scenarios~\cite{anderson2017phase,gupta2018optimized,prajapati2019polarization}, from optomechanical sensors~\cite{pooser_ultrasensitive_2015,pooser2020truncated, xia2023entanglement} to proposed dark matter detectors~\cite{caves_reframing_2020}.
More recent developments in quantum metrology have been directed towards multiple parameter estimation techniques for various applications such as imaging, microscopy, and sensor networks \cite{moreau2019imaging,ono2013entanglement,hong2021quantum,hong2022practical}. 
In the case of a network of distributed sensors, the goal of distributed quantum sensing is to measure the linear combination of spatially distributed independent parameters beyond the standard quantum limit (SQL)\cite{zhuang2018distributed,proctor2018multiparameter,guo2020distributed,zhao2021field,kim2023distributed}. Distributed quantum sensing has already been shown to provide advantages for applications such as local beam tracking and global-scale clock synchronization~\cite{qi2018ultimate,komar2014quantum}. 
Moreover, recent proposals point to further sensitivity improvements from entanglement-enhanced arrays of distributed optical sensors for phase sensing, which could enable the detection of ultra-weak signals in high energy physics, such as enhanced gravitational wave detection and direct detection of dark matter \cite{derevianko2018detecting, carney2020proposal,carney2021mechanical,attanasio2022snowmass, xia2023entanglement}.

One approach to distributed quantum phase sensing is the use of a non-linear interferometer, such as an SU(1,1) interferometer, where nonlinear amplifiers replace the  beam splitters in a standard SU(2) interferometer~\cite{hudelist2014quantum,marino2009tunable,anderson2017optimal}.  In these devices, entanglement is present inside the interferometer between the two optical paths and the nature of the quantum correlations requires the measurement of a linear combination of observables, such as the sum of the phase quadratures or the difference of the amplitude quadratures, in order to obtain the quantum noise reduction necessary to surpass the SQL. The SU(1,1) interferometer offers the potential to outperform its classical counterparts by a factor proportional to the nonlinear gain of the nonlinear amplifiers. Additionally, when seeded, the SU(1,1) interferometer can take advantage of the increased sensitivity due to the large number of photons that are used to perform the estimation~\cite{anderson2017optimal,liu2018quantum}.
A modified version of the SU(1,1) interferometer, the truncated  SU(1,1) (tSU(1,1)) interferometer, replaces the second nonlinear amplifier with two local balanced homodyne detectors, while obtaining the same quantum advantage of a full SU(1,1) interferometer~\cite{anderson2017optimal,gupta2018optimized}. Given that optical interference of the beams after sensing the parameter of interest is not needed with the tSU(1,1), it offers a natural way to implement a distributed quantum sensor to measure two distributed phases, one in each of the two entangled arms of the interferometer. In spite of this, the tSU(1,1)  has previously only been used to measure a single-phase along one of its arms~\cite{anderson2017optimal,anderson2017phase,pooser2020truncated}.  Additionally, the presence of entanglement in this system has the potential to improve the scaling of the distributed sensor when extending beyond two phases.

In this paper, we theoretically and experimentally study the use of two-mode squeezed states in a distributed quantum sensor to measure the linear combination of two phases with a quantum enhanced phase sensitivity. For the experimental two-phase distributed sensing, we achieve a quantum noise reduction of $1.7\pm0.3$ dB when using a distributed two-mode squeezed light source in a tSU(1,1) interferometer configuration. We compare the distributed system to a separable sensing approach consisting of two independent tSU(1,1) interferometers, each sensing a single-phase within the probe arm, and observe a 3 dB signal-to-noise ratio (SNR) classical improvement with respect to the separable configuration.  This allows us to experimentally demonstrate that signals hidden in the noise in the separable sensing approach become observable when measured in the distributed sensing approach with a tSU(1,1) interferometer.  We additionally extend the theory to study entanglement-enhanced distributed phase sensing for $M$ distributed phases. To this end, we consider a beam-splitter network  to extend from a two-mode entangled state to a multi-mode entangled state and show that an additional entanglement enhanced phase sensitivity can be achieved over that of an optimal distributed sensor network probed with separable quantum states. We theoretically demonstrate the expected Heisenberg sensitivity scaling with the number of sensors and show that the tSU(1,1) interferometer has the potential to surpass practical classical approaches with bright two-mode squeezed light in a distributed sensor network that enables an entanglement enhanced sensitivity beyond the SQL.

\section{Distributed quantum sensing with a truncated SU(1,1) interferometer}

\begin{figure}[b]
\centering\includegraphics[width = 12 cm]{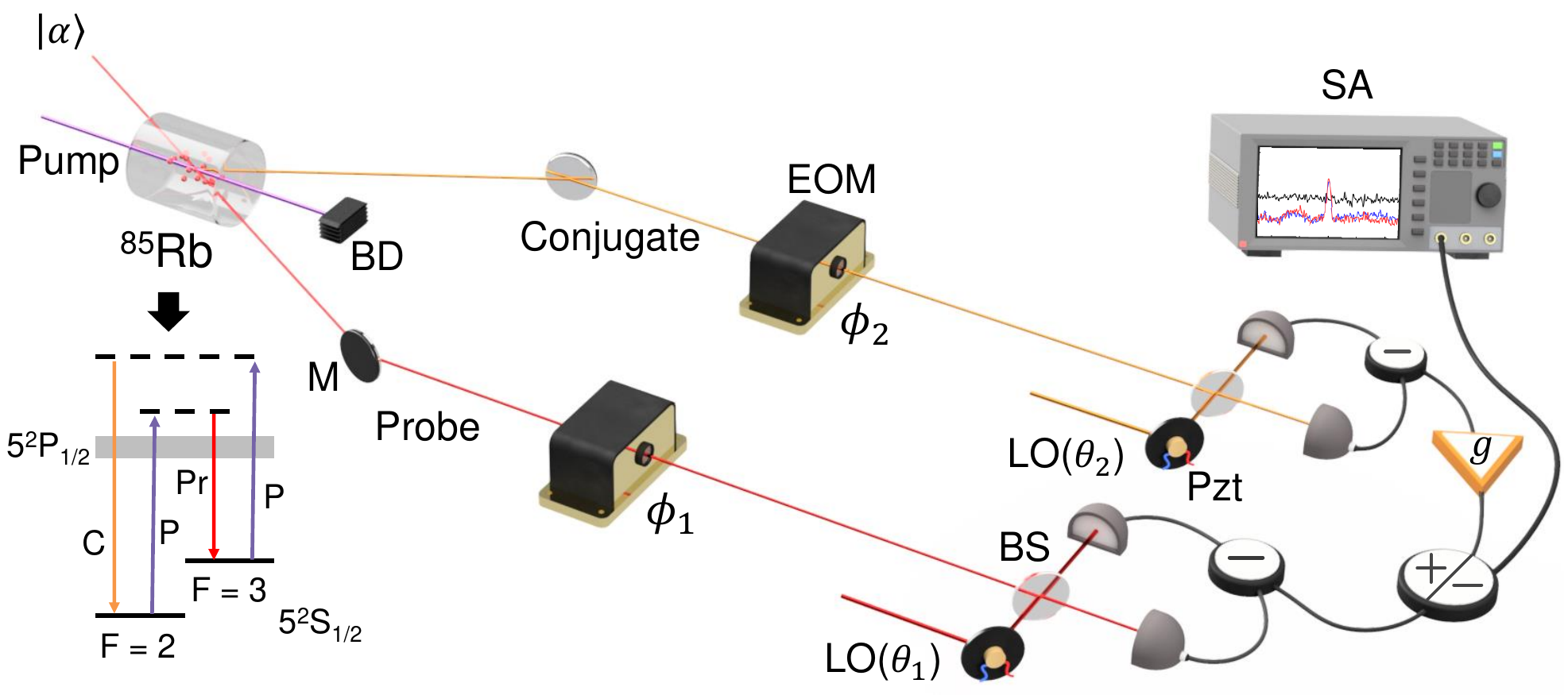}
\caption{\label{fig:1} Experimental setup to estimate the linear combination of two distributed phases in a tSU(1,1) interferometer. Bright two-mode squeezed states are generated by seeding a four-wave mixing process in a $^{85}$Rb vapor cell with a weak coherent state. Each of the modes of the two-mode squeezed state experiences a phase shift, $\phi_1$ and $\phi_2$, respectively. The two modes are then measured with two homodyne detectors to perform a joint measurement to estimate $\phi=\beta_1\phi_1+\beta_2\phi_2$. BD: beam dump; M: mirror; EOM: electro-optic modulator; LO: local oscillator; BS: beamsplitter; PZT: piezo actuator; SA: spectrum analyzer.}
\end{figure}

Our initial goal is to demonstrate a distributed quantum sensing configuration that can measure a linear combination of two distributed phases beyond the SQL with a two-mode squeezed state, as shown in the experimental schematic in Fig.~\ref{fig:1}. We start by considering two unknown phases, $\phi_1$ and $\phi_2$, that are distributed in different locations, with the the goal of estimating the linear combination  $\phi=\beta_1\phi_1+\beta_2\phi_2$, with the normalization condition $|\beta_1|+|\beta_2|=1$\cite{ge2018distributed,oh2022distributed}.
 
In general, the phase sensitivity of a measurement, $\Delta\phi$, can be evaluated from the SNR as,
\begin{eqnarray}
\text{SNR}=\frac{(\partial_{\phi}\langle\hat{X}_+\rangle)^2}{\Delta^2 X_+}\Delta^2{\phi},
\label{eq:one}
\end{eqnarray}
where $\hat{X}_+$ represents an observable, and $\Delta^2 X_+$ describes the variance of the measurement, i.e., $\Delta^2 X_+=\langle \hat{X}_+^2\rangle-\langle \hat{X}_+\rangle^2$\cite{pooser2020truncated,gupta2018optimized,anderson2017optimal,anderson2017phase}. The minimum detectable phase shift is then determined when the $\text{SNR}=1$. We thus define the limit of detection (LOD) as\cite{anderson2017phase,pai2022magneto},
\begin{eqnarray}
\Delta^2\phi=\frac{\Delta^2 X_+}{(\partial_{\phi}\langle\hat{X}_+\rangle)^2}.
\label{eq:two}
\end{eqnarray}
Here, we analyze the sensitivity (as determined by the LOD) of the distributed sensing configuration in a tSU(1,1) interferometer in the limit of bright quantum states of light, which can enable the extension of the sensing configuration to practical applications.
 
In our configuration, bright two-mode squeezed light is generated as part of the tSU(1,1) interferometer. A four-wave mixing (FWM) process in a double-lambda scheme in $^{85}$Rb vapor (see inset in Fig.~\ref{fig:1})~\cite{mccormick_strong_2007,PhysRevA.78.043816} is used to generate the required two-mode squeezed states. To generate bright beams, the probe beam for the FWM is weakly seeded with a coherent state with an average photon number $|\alpha|^2$. This then leads to the generation of an amplified probe with a photon number of $G|\alpha|^2+(G-1)$ and a corresponding conjugate beam with a photon number of $(G-1)|\alpha|^2+(G-1)$, where the $(G-1)$ term is due to spontaneous FWM while the terms proportional to $|\alpha|^2$ result from the stimulated process. In the case of  $|\alpha|^2 \gg 1$ the stimulated process dominates and the mean photon numbers can be approximated as $G|\alpha|^2$ and $(G-1)|\alpha|^2$, respectively. In these equations, $G$ is the gain of the nonlinear interaction and is related to the squeezing parameter according to $r=\text{cosh}^{-1}(\sqrt{G})$.

\begin{figure}[b]
\centering\includegraphics[width=12cm]{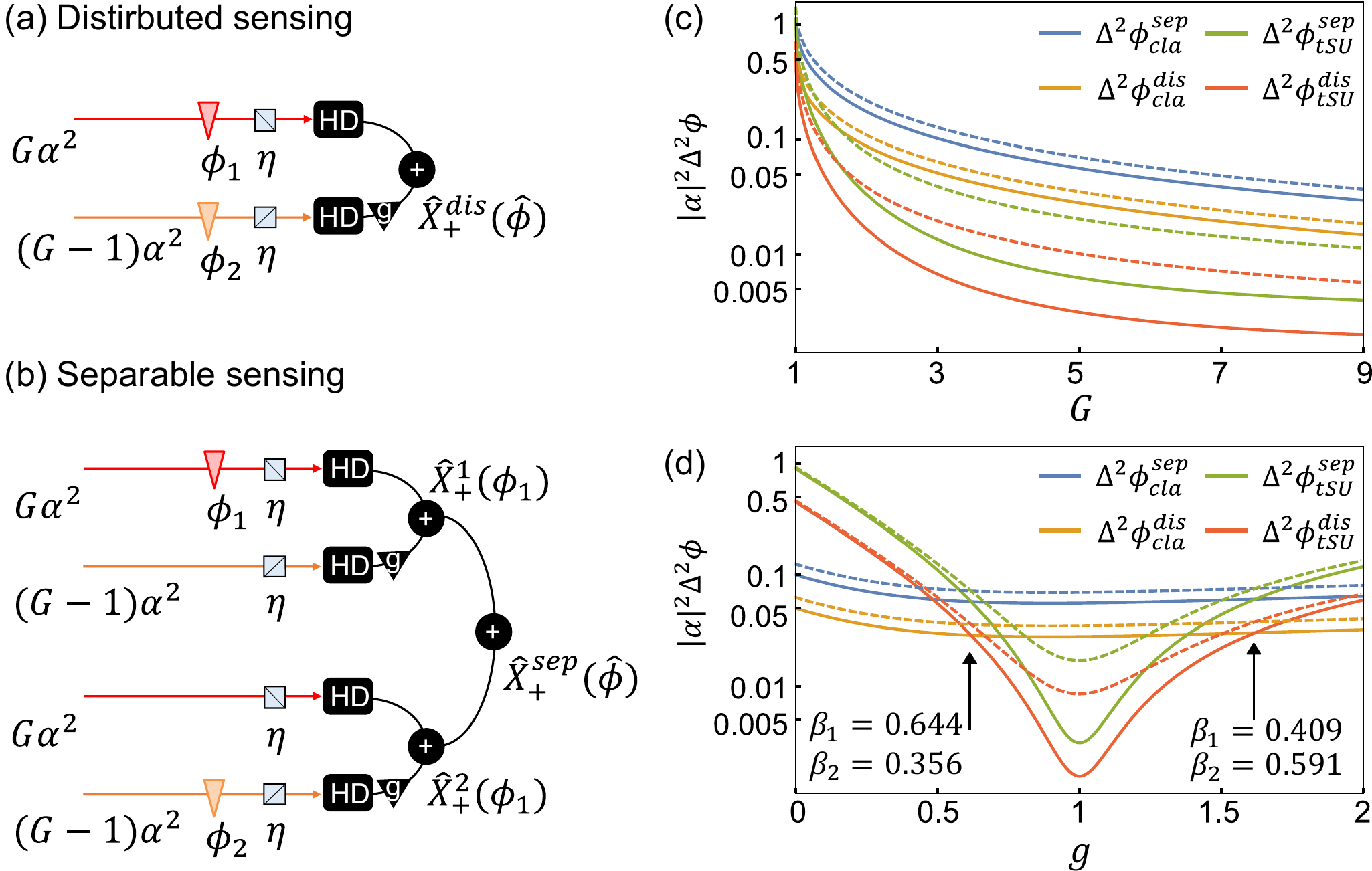}
\caption{\label{fig:2} Comparison of (a) distributed sensing and (b) separable sensing approaches. To obtain a fair comparison of the sensitivities for the two schemes and the corresponding classical configurations, the total optical power that interacts with the phase elements is held constant. (c) LOD multiplied by $|\alpha|^2$ for the two sensing approaches with quantum and classical resources as a function of gain $G$ when $|\beta_1|=|\beta_2|=1/2$. The solid traces correspond to the lossless case, with $\eta=1$, while the dashed traces correspond to the case with some loss, specifically with $\eta=0.8$, for the separable approach with coherent states (blue), distributed approach with coherent states (orange), separable approach with two-mode squeezed states (green), and distributed approach with two-mode squeezed states (red). (d) LOD multiplied by $|\alpha|^2$ for the same four cases as in (c) as a function of the classical gain factor $g$ when $G=5$ for the lossless case ($\eta=1$, solid line) and the case with losses ($\eta=0.8$, dashed line). A quantum enhanced sensitivity can be achieved from $g=0.618$ to $g=1.618$ in the lossless case, which translates to a linear phase superposition of $\phi=0.644\phi_1+0.356\phi_2$ and $\phi=0.409\phi_1+0.591\phi_2$, respectively. }
\end{figure}

After the FWM process, phase shifts $\phi_1$ and $\phi_2$ are imparted on the probe and conjugate beams, respectively. The probe and conjugate are then sent to balanced homodyne detectors with local oscillator (LO) phases $\theta_1$  and $\theta_2$, respectively, to measure the corresponding generalized quadrature operators, $\hat{X}_1(\phi_1,\theta_1)$ and $\hat{X}_2(\phi_2, \theta_2)$. To simplify the notation, we focus on the LO phases that provide the optimal LOD, specifically $\theta_1=\theta_2=\pi/2$ that correspond to measurements of the phase quadratures, and drop the $\theta$ dependence throughout the manuscript. To obtain an estimate of the linear combination of the unknown phases ($\phi$), we define the operator $\hat{X}_+$ such that $\hat{X}_+(\phi)=\hat{X}_1(\phi_1)+g\hat{X}_2(\phi_2)$, with $g$ representing a classical gain factor to scale the homodyne detection of $\hat{X}_2(\phi_2)$~\cite{woodworth2020transmission,woodworth2022transmission}. When losses are accounted for, one obtains (see Appendix)
\begin{eqnarray}
\begin{array}{rcl}
\langle\hat{X}_+(\phi)\rangle&=&2\alpha\sqrt{\eta}[\sqrt{G}\text{sin}(\phi_1)+g\sqrt{G-1}\text{sin}(\phi_2)]\\
&\approx&2\alpha\sqrt{\eta}(\sqrt{G}\phi_1+g\sqrt{G-1}\phi_2)\\
&=&2\alpha\sqrt{\eta}(\sqrt{G}+g\sqrt{G-1})\phi
\end{array}
\label{eq:three}
\end{eqnarray}
where $\eta\in[0,1]$ is the transmission through the system, such that $\eta=1$ indicates no loss. In deriving this result we make the assumption that the losses on the probe and conjugate are the same and that $\phi_1$ and $\phi_2$ are small, such that $\text{sin}\phi\approx\phi$.  Here, the weighting factors, $\beta_1$ and $\beta_2$, of the two terms in the linear combination of phases, $\phi=\beta_1\phi_1+\beta_2\phi_2$, are defined as 
$\beta_1=\sqrt{G}/(\sqrt{G}+g\sqrt{G-1})$ and $\beta_2=g\sqrt{G-1}/(\sqrt{G}+g\sqrt{G-1})$. Putting all of this together, the LOD for our distributed sensing approach with a tSU(1,1) interferometer to sense the linear combination of two phases is theoretically calculated to be
\begin{eqnarray}
\Delta^2\phi_\text{tSU}^\text{dis}=\frac{(g^2+1)(1-2\eta+2\eta G)-4g\eta\sqrt{G(G-1)}}{4|\alpha|^2\eta(\sqrt{G}+g\sqrt{G-1})^2}.
\label{eq:four}
\end{eqnarray}
We also calculate the quantum Cram\'er-Rao bound (QCRB) for our configuration to be $\Delta^2\phi_\text{QCRB}=(1-2\eta+2\eta G-2\eta\sqrt{G(G-1)})/(2|\alpha|^2\eta(\sqrt{G}+\sqrt{G-1})^2)$~\cite{woodworth2020transmission,woodworth2022transmission},  as outlined in the Appendix. As can be seen, the LOD given by  Eq.~(\ref{eq:four}) saturates the QCRB for $g=1$, which means that the measurement is determined to be optimal~\cite{you2019conclusive,woodworth2020transmission,woodworth2022transmission}.

Next, we compare the sensitivity of the distributed sensing approach in which phase shifts are simultanously present in both arms of a single tSU(1,1) interferometer with that of a separable sensing approach that consists of two tSU(1,1) interferometers each measuring a single phase along one of the arms of the interferometer, as shown in Figs.~\ref{fig:2}(a) and (b), respectively. In each case, we compare the corresponding classical configuration that is obtained when replacing the two-mode squeezed states with coherent states of the same mean number of photons. To have a fair comparison between the different configurations, we take the resources in the estimation to be the number of photons probing the phase elements, such that it is held at a constant value for all cases. The LOD for the separable sensing approach consisting of two tSU(1,1) interferometers is theoretically calculated to be $\Delta^2{\phi}_\text{tSU}^\text{sep}=2\Delta^2{\phi}_\text{tSU}^\text{dis}$ (see Appendix).  This 3~dB improvement in LOD (inverse of $\Delta^2{\phi}$) of the distributed over the separable sensing approach is a classical effect that results from the doubling of the signal size in the distributed approach due to all the resources being used to probe the phase shifts, as opposed to having only a single arm of the interferometer probe a single phase shift.  Finally, the LODs for the  distributed and separable approaches with classical light can be shown to be given by $\Delta^2{\phi}_\text{cla}^\text{dis}=(1+g^2)/(4 |\alpha|^2 \eta (\sqrt{G}+g\sqrt{G-1}))$ and $\Delta^2{\phi}_\text{cla}^\text{sep}=2\Delta^2{\phi}_\text{cla}^\text{sep}$, respectively. Note that the same 3~dB improvement in LOD of the distributed over the separable approach is obtained with classical resources.

Figure~\ref{fig:2}(c) shows the theoretically calculated LODs multiplied by $|\alpha|^2$ as a function of gain ($G$) with $\eta=1$ (solid line) and $\eta=0.8$ (dashed line) when $\beta_1=\beta_2=1/2$. There is a clear quantum enhanced sensitivity for the distributed two phase sensing with a tSU(1,1) interferometer over that of the corresponding classical approach.  As can be seen, the distributed approach reaches the expected classical 3 dB advantage over that of the separable sensing approach for large gains. In addition, we calculate the LODs as a function of the classical gain factor $g$ when $G=5$, as shown in Fig.~\ref{fig:2}(d). A quantum enhanced sensitivity over that of the corresponding classical benchmark can be obtained in the range of $0.618<g<1.618$, which translates to an adjustment of $\phi$ from $\phi=0.644\phi_1+0.356\phi_2$ to $\phi=0.409\phi_1+0.591\phi_2$. Note that the weightings between phases $\phi_1$ and $\phi_2$ are adjustable while still maintaining a certain amount of quantum noise reduction in the sensor even without optimizing $g$. Further details on the sensitivity calculations for both the distributed and separable approaches can be found in the Appendix.

\section{Experimental results}\label{experiment}

The tSU(1,1) interferometer used for our distributed sensing experiments uses a two-mode squeezed light  state generated with a FWM process in a $^{85}\text{Rb}$ vapor cell (see Appendix for details) to probe two independent phase shifts, as shown in Fig~\ref{fig:1}.  As noted previously, this configuration can be used for distributed sensing as there is no need to optically interfere the two modes after interfacing with the phase elements. For our experiment, a strong (360~mW) pump beam and a weak (5~$\mu$W) probe seed beam (red-shifted by 3.044~GHz from the frequency of the pump), are mixed at an angle of 0.3$^\circ$ in a 12.7~mm long $^{85}\text{Rb}$ vapor cell, which is heated to 100$^\circ \text{C}$.  The pump and probe have beam diameters of 920~$\mu$m and 560~$\mu$m, respectively.  This configuration leads to a gain of  $G\approx 5$, such that seeding the probe beam leads to an optical power for the probe and conjugate beams of 26~$\mu$W (an average of $10\times10^{13}$ photons) and 17~$\mu$W (an average of $6.8\times10^{13}$ photons), respectively. These power levels are kept constant for all experiments and used to probe the phases elements.  Note that the generated bright beams of light have significantly larger optical power than squeezed vacuum states or Fock states, and thus they can be applied to practical sensing applications requiring few tens of microwatts\cite{guo2020distributed,hong2021quantum,lawrie2019quantum}. 

To generate the LOs required for the balanced homodyne detectors, we use a second independent FWM process (not shown in Fig.~\ref{fig:1}) as shown in the Appendix.  For this, a coherent state with 44.5~$\mu$W is also injected into the same vapor cell and mixed with its own pump beam in order to generate an LO pair with matching spatial mode properties to those of the probe and conjugate beams~\cite{boyer2008}.  The LOs also experience gain in the cell, leading to optical powers approximately an order of magnitude larger than their respective probe and conjugate beams due to the larger seed power. For homodyne detection, the probe and conjugate are interfered with their respective LOs, after which they are independently detected with a difference measurement using a pair of balanced photodiodes.  The external phase references are set by the LOs, such that the probe and conjugate are locked with a $\pi/2$ phase offset to these references to measure phase squeezing (for details on the LO lock see the Appendix). The observable $\hat{X}_+(\phi)$ is then measured using a hybrid junction to obtain the sum of the signals from the dual balanced homodyne detectors and read out with a spectrum analyzer (SA). 

Before detection, phase shifts at a frequency of 300 kHz are imparted on the two mode squeezed state using either one, or two, electro-optic modulators (EOMs).  An EOM is placed in the paths of the probe and conjugate beams for all the measurements, even if they are not actively imparting a phase shift, as shown in Fig~\ref{fig:1}. The signal resulting from the 300 kHz modulation appears as a peak on the SA. The shot noise limit (SNL) for all signals is measured with the same setup when only the order of magnitude larger LOs are incident on the balanced detectors and the squeezed light is blocked.

In the experiment, we set the phases along each arm of the interferometer to the same value $\phi=\phi_1=\phi_2$, and $\beta_1$ and $\beta_2$ are set to $1/2$ by adjusting the classical gain $g$. For the separable sensing approach only one phase element is active at a time and the signal is  obtained by summing the data acquired with only $\phi_1$ or $\phi_2$ active, which is equivalent to the separable sensing scheme shown in Fig.~\ref{fig:2}(b). For the distributed sensing, on the other hand, we apply both phase shifts simultaneously along each optical path, as shown in Fig.~\ref{fig:2}(a), with the phase shifts equal to those applied in the two measurements for the separable sensing approach. To obtain the SNL for each of the approaches, we use the equivalent configuration with the two-mode squeezed state replaced with conherent states. For all measurments, the optical power along the probe and conjugate arms are held constant. 

\begin{figure}[t]
\centering\includegraphics[width=12cm]{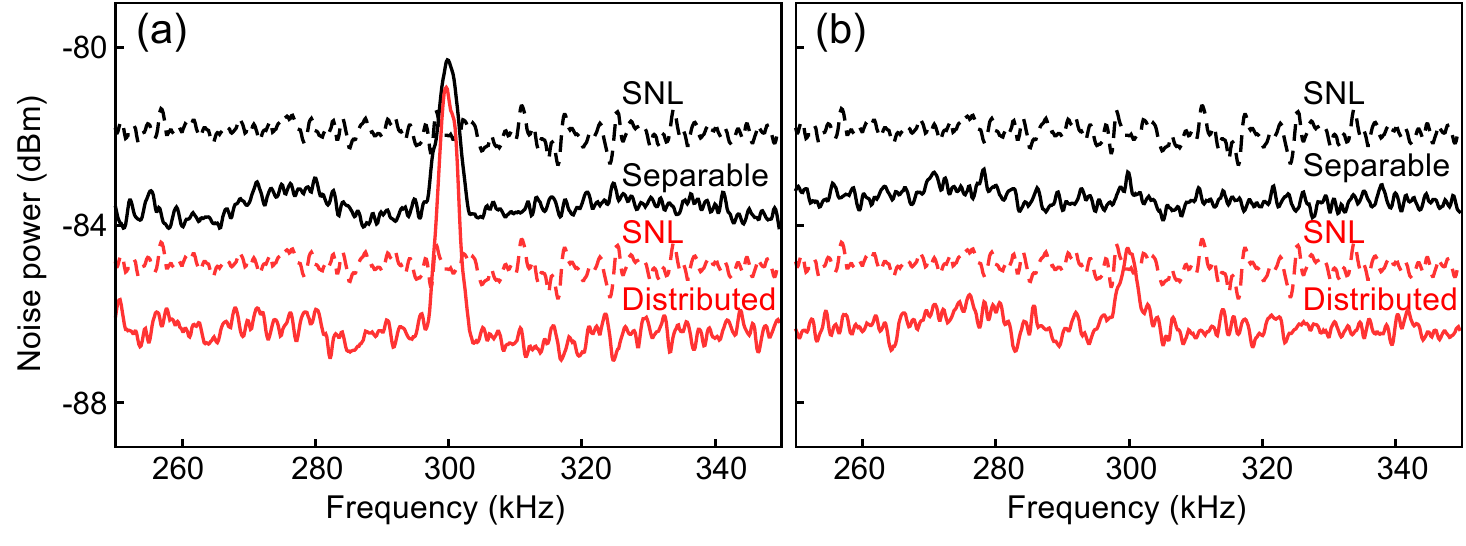}
\caption{\label{fig:3} Measured spectrum analyzer traces when driving the EOM(s) with a signal amplitude of (a) 30 mV and (b) 10 mV at 300 kHz. The black (red) solid and dashed lines indicate the separable (distributed) sensing approach and the corresponding SNL, respectively. (a) A quantum noise reduction of $1.7\pm0.3$ dB with respect to the SNL is achieved for both the separable and distributed sensing cases. Additionally, the SNR is classically improved by $3.0\pm0.5$ dB when using the distributed sensing approach as compared to the separable sensing approach. (b) For a small enough modulation strength, the signal can no longer be resolved with the separable sensing approach, whereas for the distributed sensing measurements the signal is resolvable. Settings for the spectrum analyzer: resolution bandwidth (RBW) = 3 kHz; video bandwidth (VBW) = 100 Hz; sweep time = 1 s. All traces are averaged 15 times.}
\end{figure}

The measured signals are shown in Fig.~\ref{fig:3}(a), with the black traces corresponding to the separable approach with two-mode squeezed states (solid traces) and coherent states (dashed traces) and the red traces corresponding to the distributed approach with two-mode squeezed states (solid traces) and coherent states (dashed traces). As can be seen for the separable approach, a $1.7\pm0.3$ dB quantum enhancement with respect to the corresponding SNL was achieved when using the two-mode squeezed light source. This degree of quantum enhancement is maintained in the distributed sensing approach, which shows the usefulness of the tSU(1,1) for distributed sensing. In addition, when comparing the distributed and separable sensing approaches, we observe a classical SNR improvement  of $3.0\pm0.5$ dB when using either quantum or classical resources, as expected, as a result of the constraint of keeping the total optical power constant (i.e. total  optical power of the probe and conjugate beams in an interferometer). A description of the SNR data analysis is provided in the Appendix.

 We further demonstrate that the classical enhacement makes it possible for the distributed sensing approach to resolve smaller modulations than those possible with the separable sensing approach. For this, we impose smaller phase shifts $\phi_1$ and $\phi_2$.  In the separable sensing approach, for which these  phase shifts are applied one at a time, the total phase shift is not resolvable, as shown by the black trace in Fig.~\ref{fig:3}(b). However, for the distributed sensing approach, when the optical power is held constant and both phase shifts are simultaneously applied, it is possible to resolve the small modulation, as can be seen from the red solid line in Fig.~\ref{fig:3}(b). These results demonstrate, in practice, that distributed sensing in a tSU(1,1) interferometer enables the measurement of smaller phase modulations than a separable sensing approach consisting of two tSU(1,1) interferometers or with the corresponding classical interferometer alone.

\section{Distributed multi-phase sensing with a tSU(1,1) interferometer}

We now extend the theory developed for the case of two distributed phases with a tSU(1,1) interferometer to that of a generalized configuration for sensing $M$ distributed phases with $M$ entangled probe beams in a tSU(1,1) interferometer. The two-mode squeezed states that were the basis for our experiments can also be used to generate a multi-mode entangled state by utilizing a balanced beamsplitter network (BSN) consisting of beamsplitter arrays, as proposed in Refs.~\cite{zhuang2018distributed,guo2020distributed}. The multi-mode entangled state can then be used to measure a linear combination of $M$ phases.  Here, we consider the average of the phases, i.e. $\phi=\sum_{j=1}^{M}\phi_j/M$, with $M$  being the number of unknown phases.  

\begin{figure}[htbp]
\centering\includegraphics[width=11cm]{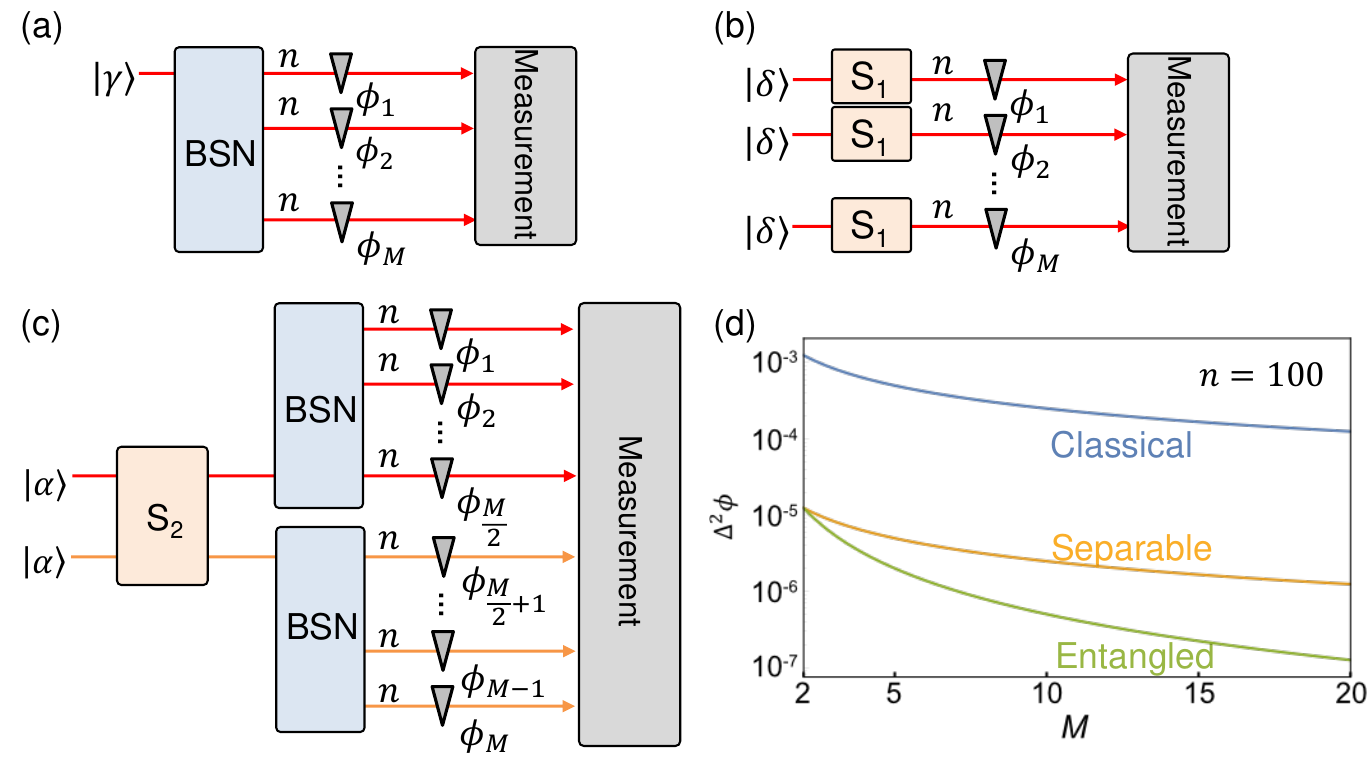}
\caption{\label{fig:5} Schemes for distributed sensing of a linear superposition of $M$ unknown phases with (a) a classical sensing scheme, (b) a separable scheme composed of multiple independent single-mode squeezed states, and (c) a multi-mode entangled scheme based on a two-mode squeezed state in a tSU(1,1) interferometer. The average photon number $n$, which is incident on each phase elements, is kept fixed for all the schemes. (d) Theoretical prediction for the LODs, in the case of no loss, as a function of the number of unknown phases ($M$) for $n=100$. Blue, orange, and green solid lines indicate the LOD for the classical, separable, and entangled schemes, respectively. As can be seen, the entangled scheme has a better sensitivity than the other schemes for the estimation of the average of the distributed phases. 
$S_1$: single-mode squeezing operator; $S_2$: two-mode squeezing operator; BSN: beamsplitter network.}
\end{figure}

We compare the LODs for the measurement of the average of $M$ unknown phases for the three cases shown in Figs.~\ref{fig:5}(a) through~\ref{fig:5}(c): a classical scheme, a separable scheme composed of multiple independent single-mode squeezed states~\cite{guo2020distributed}, and the multi-mode entangled scheme with the two-mode squeezed states in a tSU(1,1) interferometer, respectively. Note that the separable scheme we use for this comparison is different from the one used in Sect.~\ref{experiment} for the experimental implementation with two distributed phases. For the experimental implementation we considered  the equivalent separable configuration based on two-mode squeezed states, as it can be directly implemented with the same setup to minimize any artifacts, such as changes in alignment, that can lead to artificial changes in the signal size. For the theoretical analysis we consider instead the optimal separable configuration, as shown in~\cite{guo2020distributed}, to better illustrate the quantum advantage from entanglement of the proposed scheme. Finally, we constrain the average photon number, $n$, incident on each phase element to be constant for all three schemes. This is done by appropriately setting the input seed powers for the three cases (i.e. $|\gamma|^2$, $|\delta|^2$, and $|\alpha|^2$). 

More specifically, the classical scheme consists of a coherent state as the input to a BSN, which is used to distribute it to $M$ phases elements that impart phase shifts $\phi_1$ to $\phi_M$ prior to a joint measurement. The separable scheme consists of $M$ single-mode squeezed states that independently probe $M$ distributed phases prior to a joint measurement. Finally, for the $M$ mode entangled case, BSNs are used to distribute a two-mode squeezed states to probe teh $M$ distributed phase elements. In order to balance the power equally among the $M$ phase elements, we consider the case in which both probe and conjugate beams of the tSU(1,1) are weakly seeded with a coherent state with mean photon number $|\alpha|^2$, such that the total number of photons prior to the BSNs is equal to $N_\text{tot}=2(G-1)+4G|\alpha|^2+2|\alpha|^2\left(-1+2\sqrt{G(G-1)}\right)$, where $G$ is the nonlinear gain. The BSNs split the two-mode squeezed state into $2d = M$ modes, such that the average photon number probing  each phase element is equal to $n = N_\text{tot}/M$  prior to detection with a joint measurement.  

For all three schemes, we calculate the LOD when all phase elements introduce the same phase shift, $\phi_{1}=\phi_{2}=...=\phi_{M}=\phi$, as a function of both $n$ and $M$ for the lossless case with  $\eta=1$. Given this, we find that the optimal sensitivities for each scheme takes the form 
\begin{eqnarray}
\Delta^2\phi_\text{cl}=\frac{1}{4Mn},\\
\Delta^2\phi_\text{separable}=\frac{1}{4Mn(n+1)},\\
\Delta^2\phi_\text{entangled}=\frac{1}{2Mn(Mn+2)}.
\label{eq:5}
\end{eqnarray}
Full details of the complete calculation for the multi-mode entangled scheme for distributed sensing in a tSU(1,1) interferometer can be found in the Appendix, along with the required calculations for determining the LOD for the other two schemes.

Notably, we  find that while the separable case with quantum resources achieves a Heisenberg scaling with the number of photons, the entangled scheme with a tSU(1,1) interferometer achieves Heisenberg scaling with both the average photon number $n$ and the number of modes $M$.  Figure~\ref{fig:5}(d) shows the LODs for all three cases as a function of the number of unknown phases $M$ for the lossless case with an average photon number $n=100$. Both schemes with quantum resources show a significant improvement in their LOD in comparison to the classical scheme. Additionally, when more than two phases are sensed in the distributed sensor network, there is a clear improvement in sensitivity when using the entangled scheme in a tSU(1,1) interferometer over that of using the separable scheme, thus demonstrating that an additional entanglement enhanced sensitivity is possible. An extension of this work, beyond the scope of this paper, would be to study the quantum enhancement in distributed phase sensing using different entanglement resource states beyond those shown here. These entangled resources could include states such as the dual-rail cluster states, which can be generated with an optical spatial mode comb ~\cite{pooser2014continuous}, and could also be used in distributed phase sensing.

\section{Conclusion}
We theoretically and experimentally investigate the measurement of a linear combination of distributed phases in a sensor network using bright states of light. For the distributed sensing of two phases, both theory and experiment show that distributed phase sensing in a nonlinear interferometer provides a limit of detection improvement over the cases of the corresponding separable sensing and classical sensing approaches. In the experimental demonstration, the distributed sensing approach in a tSU(1,1) interferometer provides up to 1.7 dB of quantum noise reduction compared to the classical approach and a classical 3 dB improvement in SNR compared to the equivlaent separable sensing approach based on two-mode squeezed states. Further, we showed that the distributed phase sensing scheme in a tSU(1,1) interferometer can be generalized to estimate the linear superposition of multiple phases with multi-mode entangled states. This generalized $M$ mode entangled distributed phase sensor can be built by starting from the bright two-mode squeezed source used in our experiments. This approach provides an improved sensitivity scaling over that of the optimal separable distributed sensor using independent single-mode squeezed states of light. Our results pave the way for developing quantum enhanced sensor networks that can provide a quantum advantage that also leverages entanglement for a range of sensing applications from global-scale clock synchronization to high energy physics, including cases that benefit from bright probing fields to obtain sensitivities that can go beyond the classical state-of-the art.

\section*{Acknowledgments}
SH led the experimental effort and data analysis, RCP oversaw the research, MTF provided advised on the experiment, MAF assisted in building the experiment; they were supported by the U.S. DOE Office of Science, Office of High Energy Physics, QuantISED program (under FWP ERKAP63). AMM oversaw the research and assisted with theory and data analysis, and CEM advised and assisted with data collection and experimental design; they received support from the U.S. DOE, Office of Science, National Quantum Information Science Research Centers, Quantum Science Center. SH and CL were supported by an Institute for Information \& Communications Technology Planning \& Evaluation (IITP) grant funded by the Korea government (MSIT) (RS-2023-00222863), and SH was also supported by the National Research Foundation of Korea (NRF) (RS-2023-00283146) for writing of the manuscript. All authors contributed to the analysis and writing.

\section*{Appendix}

\section*{Phase sensitivity Calculations}
Here we derive the phase sensitivities for the separable and distributed two-phase sensing approaches considered for the experimental implementation in the main text.

\subsection*{Distributed sensing configuration}\label{sec:distsens}
\begin{figure}[b]
\centering\includegraphics[width=13.5cm]{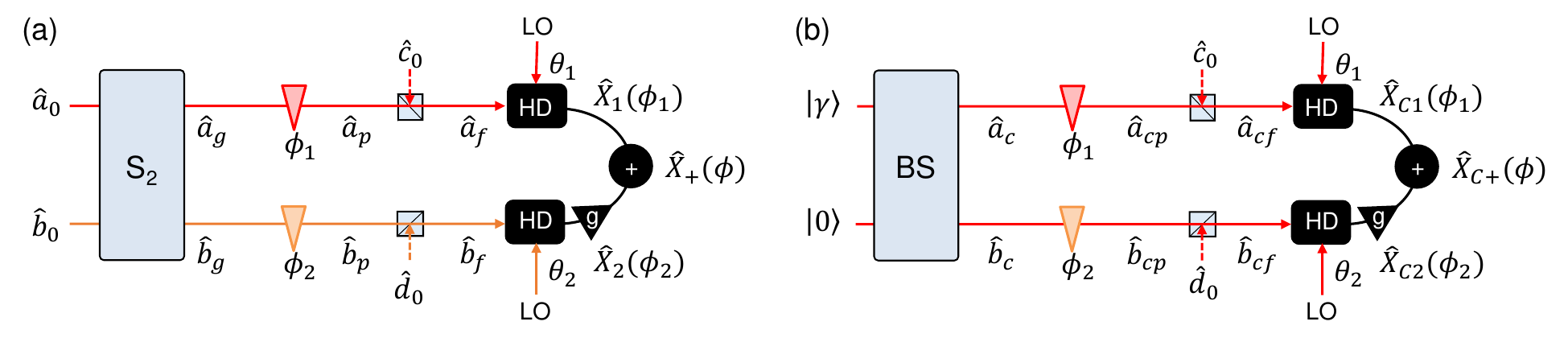}
\caption{\label{fig:s1} Schematics for the models used to derive the phase sensitivity when using (a) a two-mode squeezed state in a tSU(1,1) interferometer and (b) the corresponding classical configuration. S2: two-mode squeezing operator; BS: beam splitter.}
\end{figure}

First, we consider the use of a truncated SU(1,1) interferometer (tSU(1,1)) as a distributed sensor, as shown schematically in Fig.~\ref{fig:s1}(a). For the generation of the two-mode squeezed state, we consider two modes, $\hat{a}_0$ and $\hat{b}_0$, as the input for the four-wave mixing (FMW) process that takes place in the $^{85}$Rb vapor cell. To model the phase insensitive configuration used in the experiment, mode $\hat{a}_0$ is taken to be a coherent state used as a weak probe seed, with $\langle \hat{a}^\dagger_0\hat{a}_0\rangle=|\alpha|^2$, while mode $\hat{b}_0$ is take to be a vacuum state, i.e. $\langle \hat{b}^\dagger_0\hat{b}_0\rangle=0$, as the conjugate is unseeded. In the ideal case, the FMW process in the $^{85}$Rb vapor cell leads to a nonlinear gain, $G=\cosh^2(r)$ where $r$ is the squeezeing paramter, that transforms the input modes into output modes $\hat{a}_g$ and $\hat{b}_g$ for the probe and conjugate beams, respectively, according to~\cite{jasperse2011relative}:
\begin{equation}
\begin{pmatrix} 
\sqrt{G} & 0 & 0 & \sqrt{G-1} \\ 
0 & \sqrt{G} & \sqrt{G-1} & 0 \\ 
G & \sqrt{G-1} & \sqrt{G} & 0 \\ 
\sqrt{G-1} & 0 & 0 & \sqrt{G} 
\end{pmatrix}\cdot
\begin{pmatrix} 
\hat{a}_0 \\ 
\hat{a}^\dagger_0 \\ 
\hat{b}_0 \\ 
\hat{b}^\dagger_0 
\end{pmatrix}=
\begin{pmatrix} 
\hat{a}_g \\ 
\hat{a}^\dagger_g \\ 
\hat{b}_g \\ 
\hat{b}^\dagger_g 
\end{pmatrix}.
\label{eq:amplifying}
\end{equation}
After the FMW process, modes $\hat{a}_g$ and $\hat{b}_g$ interact with the phase elements to introduce phase shifts $\phi_1$ and $\phi_2$, respectively, to produce output modes $\hat{a}_p=e^{-i\phi_1}\hat{a}_g$ and $\hat{b}_p=e^{-i\phi_2}\hat{b}_g$. Additionally, optical losses in the paths of the beams need to be considered, which leads to the transformations  $\hat{a}_f=\sqrt{\eta}\hat{a}_p+\sqrt{1-\eta}\hat{c}_0$ and $\hat{b}_f=\sqrt{\eta}\hat{b}_p+\sqrt{1-\eta}\hat{d}_0$, where we have considered the case of equal loss for both probe and conjugate beams. In this case, we model the loss with a beam splitter transformation with transmission $\eta$ (where ${\eta=1}$ is the lossless case) for each of the beams with vacuum modes $\hat{c}_0$ and $\hat{d}_0$ coupling in through the unused input port of the beam splitters in the probe and conjugate paths, respectively. Given that the FWM process is seeded with $|\alpha|^2\gg 1$, the optical power of the probe and conjugate can be approximated as $\langle \hat{a}^\dagger_f\hat{a}_f\rangle=\eta(-1+G+G|\alpha|^2)\approx \eta G|\alpha|^2$ and $\langle \hat{b}^\dagger_f\hat{b}_f\rangle=\eta(-1+G+(G-1)|\alpha|^2)\approx \eta (G-1)|\alpha|^2$, respectively.

The modes after probing the phase elements, $\hat{a}_f$ and $\hat{b}_f$, are measured with two independent balanced homodyne detectors that have local oscillators (LOs) with phases $\theta_1$ and $\theta_2$, respectively, that serve as external phase references. We define generalized quadrature operators $\hat{X}_1(\phi_1,\theta_1)=e^{-i\theta_1}\hat{a}^\dagger_f+e^{i\theta_1}\hat{a}_f$ and $\hat{X}_2(\phi_2,\theta_2)=e^{-i\theta_2}\hat{b}^\dagger_f+e^{i\theta_2}\hat{b}_f$ for the probe and conjugate, respectively. These genrealized quadratures are combined to obtain the joint quadrature operator $\hat{X}_+(\phi)=\hat{X}_1(\phi_1,\theta_1=\pi/2)+g\hat{X}_2(\phi_2,\theta_1=\pi/2)$, where $g$ is a classical gain factor that provides a relative scaling between the probe and conjugate homodyne detectors and the external phases references given by the LOs are set to $\theta_1=\theta_2=\pi/2$ to perform measurements of the phase quadratures. In this case, the expectation value and variance for $\hat{X}_+$ are given by
\begin{eqnarray}
\langle \hat{X}_+(\phi) \rangle & = &2\alpha\sqrt{\eta}[\sqrt{G} \text{ sin}(\phi_1)+g\sqrt{G-1}\text{ sin}(\phi_2)] \nonumber\\
& \approx & 2\alpha\sqrt{\eta}(\sqrt{G} \phi_1+g\sqrt{G-1}\phi_2) \nonumber\\
& = & 2\alpha\sqrt{\eta}(\sqrt{G} +g\sqrt{G-1})\phi,\label{eq:exp}\\
\langle \Delta^2\hat{X}_+ \rangle & = & \langle \hat{X}^2_+(\phi) \rangle-\langle \hat{X}_+(\phi) \rangle^2 \nonumber\\
& = &(g^2+1)(1-2\eta+2\eta G)-4g\eta\sqrt{G(G-1)}\text{cos}(\phi_1+\phi2) \nonumber\\
& \approx & (g^2+1)(1-2\eta+2\eta G)-4g\eta\sqrt{G(G-1)}, \label{eq:var}
\end{eqnarray}
where the linear combination of the two measured phases is defined as $\phi=\beta_1\phi_1+\beta_2\phi_2$, with $\beta_1=\sqrt{G}/(\sqrt{G}+g\sqrt{G-1})$ and $\beta_2=g\sqrt{G-1}/(\sqrt{G}+g\sqrt{G-1})$ such that the normalization condition $|\beta_1|+|\beta_2|=1$ is satisfied. We also assume that the phase shifts $\phi_1$ and $\phi_2$ are small enough such that the small-angle approximations $\text{sin}(\phi)\approx\phi$ and $\text{cos}(\phi_1+\phi_2)\approx1$ can be used. Using Eqs.~(\ref{eq:exp}) and Eq.~(\ref{eq:var}), one can obtain the limit of detection (LOD) for distributed sensing in a tSU(1,1) interferometer as follows:
\begin{eqnarray}
\Delta^2\phi_\text{tSU}^\text{dis}=\frac{\langle\Delta^2\hat{X}_+\rangle}{(\partial_{\phi}\langle\hat{X}_+\rangle)^2}=\frac{(g^2+1)(1-2\eta+2\eta G)-4g\eta\sqrt{G(G-1)}}{4|\alpha|^2\eta(\sqrt{G}+g\sqrt{G-1})^2}.
\label{eq:LODdtSU}
\end{eqnarray}

We can compare this result to the fundamental limit given by the quantum Cram{\'e}r-Rao bound (QCRB) for a two-mode squeezed state in a tSU(1,1) interferometer.  Given that the two-mode squeeze states is a Gaussian quantum state, it is completely characterized by displacement vector $\vec{d}$ and covariance matrix $\sigma$~\cite{vsafranek2018estimation}. The elements of the displacement vector $\vec{d}$ are defined according to $d_i=\langle \hat{A}_i \rangle$, while the ones for the covariance matrix are defined as  $\sigma_{i,j}=\langle \hat{A}_i\hat{A}^\dagger_j+\hat{A}^\dagger_j\hat{A}_i \rangle-2\langle \hat{A}_i \rangle\langle \hat{A}^\dagger_j \rangle$, where $\hat{A}=(\hat{a}_f,\hat{b}_f,\hat{a}^\dagger_f,\hat{b}^\dagger_f)^T$. Note that $\hat{a}_f$ and $\hat{b}_f$ are the modes after the phase elements and losses in Fig.~\ref{fig:s1}(a). In the limit of bright beams, the quantum Fisher information (QFI) takes the form~\cite{vsafranek2018estimation, woodworth2020transmission}
\begin{equation}
F_{i,j}=2\partial_{\phi_i}\vec{d}^\dagger \sigma^{-1} \partial_{\phi_j}\vec{d}.
\label{eq:QFI}
\end{equation}
Using the QFI, one can show that the QCRB for phase estimation is given by
\begin{equation}
\Delta^2\phi_\text{QCRB}=\begin{pmatrix} 
\beta_1 & \beta_2
\end{pmatrix}\cdot\vec{F}^{-1}\cdot
\begin{pmatrix} 
\beta_1 \\
\beta_2
\end{pmatrix}=\frac{1-2\eta+2\eta G-2\eta\sqrt{G(G-1)}}{2|\alpha|^2\eta(\sqrt{G}+\sqrt{G-1})^2}.
\label{eq:QCRB}
\end{equation}
For here we can see that $\Delta^2\phi_\text{QCRB}=\Delta^2\phi_\text{tSU}^\text{dis}$ for $g=1$. Note that this is the case given that equal losses are considered for the probe and conjugate beams. In this case, the phase sensitivity of the tSU(1,1) interferometer saturates the QCRB.

Next, we calculate the phase sensitivity for the corresponding classical configuration using two coherent states, $\hat{a}_c$ and $\hat{b}_c$, to probe the phases elements $\phi_{1}$ and $\phi_{2}$, respectively, as shown in Fig.~\ref{fig:s1}(b). The coherent states have the same optical power, $\langle \hat{a}_c^\dagger\hat{a}_c\rangle=G|\alpha|^2$ and $\langle \hat{b}_c^\dagger\hat{b}_c\rangle=(G-1)|\alpha|^2$, as the probe and conjugate in the tSU(1,1) interferometer in the limit of $|\alpha|^2\gg1$. Experimentally, an input coherent state $|\gamma\rangle$ is split into two with a beamsplitter to generate beams with the required optical powers. After the phase elements, optical losses are considered with beamsplitter transformations, which leads to $\langle \hat{a}_{cf}^\dagger\hat{a}_{cf}\rangle=\eta G|\alpha|^2$ and $\langle \hat{b}_{cf}^\dagger\hat{b}_{cf}\rangle=\eta (G-1)|\alpha|^2$, respectively. Following the same derivation as the one used to obtain $\Delta^2\phi_\text{tSU}^\text{dis}$, one can show that the expectation value for the joint quadrature operator for the classical distributed sensing case is given by $\langle \hat{X}_{C+}(\phi)\rangle=2\alpha\sqrt{\eta}(\sqrt{G}+g\sqrt{G-1})\phi$, where $\phi=\beta_1\phi_1+\beta_2\phi_2$ with the same definition for $\beta_1$ and $\beta_2$ as used for the tSU(1,1) interferometer. Similarly, the variance can be calculated  to be $\langle\Delta^2\hat{X}_{C+}\rangle=(g^2+1)$, which leads to a LOD for the classical distributed sensing of the form
\begin{eqnarray}
\Delta^2\phi_\text{cla}^\text{dis}=\frac{g^2+1}{4|\alpha|^2\eta(\sqrt{G}+g\sqrt{G-1})^2}.
\label{eq:cldisLOD}
\end{eqnarray}
Equation~(\ref{eq:cldisLOD}) serves as the classical limit to compare with for distributed sensing with the tSU(1,1) interferometer in order to demonstrate a quantum-enhanced sensitivity.

\subsection*{Separable sensing configuration}

\begin{figure}[b]
\centering\includegraphics[width=13cm]{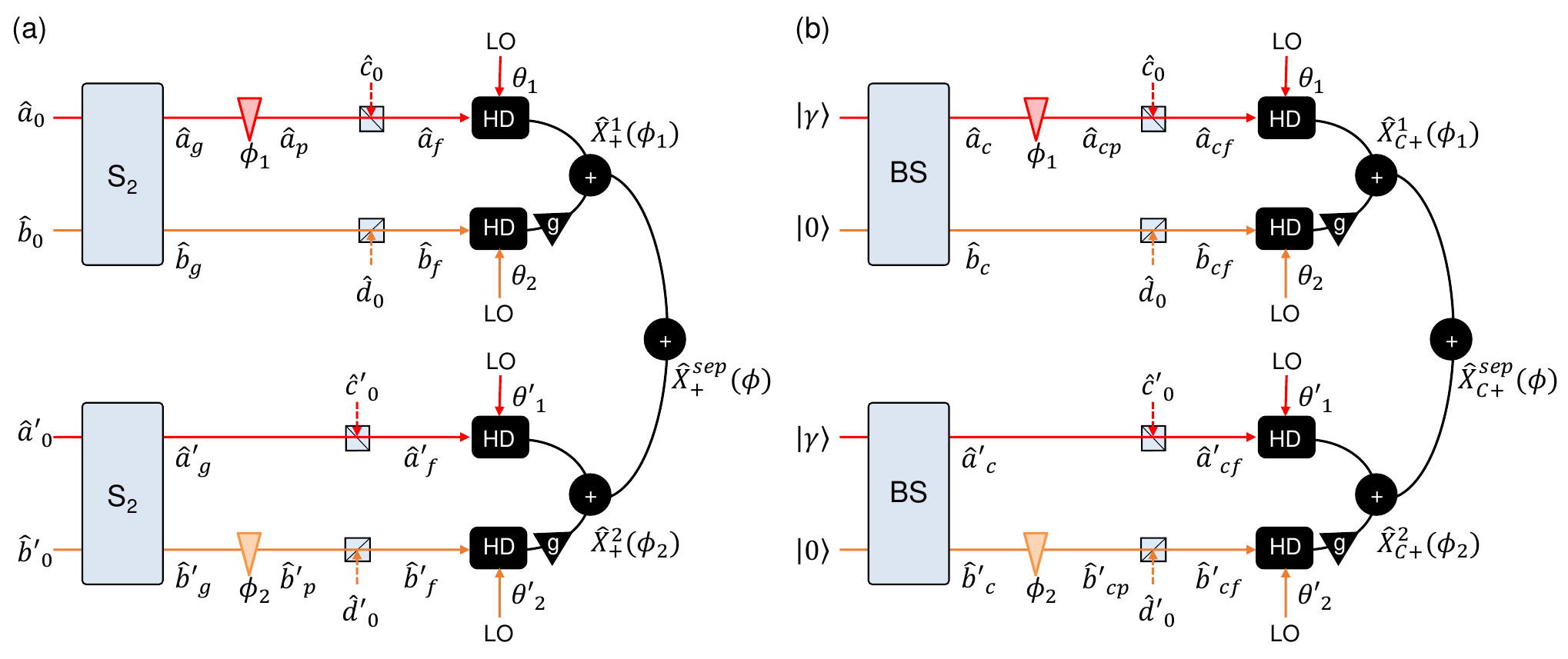}
\caption{\label{fig:s2} Schematics for the models considered for the separable sensing configurations when using (a) two tSU(1,1) interferometers and (b) the corresponding classical configuration. The separable sensing configurations that we consider are composed of two interferometers with each one measuring a single phase. S2: two-mode squeezing operator; BS: beam splitter.}
\end{figure}

For the separable sensing configuration with quantum resources, we consider the case of two tSU(1,1) interferometers with each one measuring a single phase, either $\phi_{1}$ or $\phi_{2}$, as shown in  Fig.~\ref{fig:s2}(a).  To calculate the phase sensitivity in this case, we can take the results of Sect.~\ref{sec:distsens} and consider the cases when either one phase or the other are set to zero.  For example, for the tSU(1,1) interferometer for which only mode $\hat{a}_g$ experiences a phase shift of $\phi_1$, the two output modes would be given by $\hat{a}_p=e^{-i\phi_1}\hat{a}_g$ and $\hat{b}_p=\hat{b}_g$. In this case the generalized measured quadrature takes the form  $\hat{X}^1_+({\phi_1})=(-i\hat{a}^\dagger_f+i\hat{a}_f)+g(-i\hat{b}^\dagger_f+i\hat{b}_f)$, which depends only on $\phi_1$ through $\hat{a}^\dagger_f$ and $\hat{a}_f$. In the same way, the second tSU(1,1) will lead to a generalized measured quadrature of the form  $\hat{X}^2_+({\phi_2})=(-i\hat{a}'^\dagger_f+i\hat{a}'_f)+g(-i\hat{b}'^\dagger_f+i\hat{b}'_f)$, which depends only on $\phi_2$ through  $\hat{b}'^\dagger_f$ and $\hat{b}'_f$. For $\hat{a}^\dagger_f$ and $\hat{a}_f$ both cases we have already assumed the phases of the LOs to be set to $\pi/2$. To obtain a measure of the linear combination of the phases, $\phi$, we define a joint quadrature operator of the form $\hat{X}^\text{sep}_+({\phi})=\hat{X}^{1}_+({\phi_1})+\hat{X}^2_+({\phi_2})$. Through the use of Eqs.~(\ref{eq:exp}) and~(\ref{eq:var}) we can then show that the expectation value and variance of $\hat{X}^\text{sep}_+({\phi})$ take the form
\begin{eqnarray}
\langle \hat{X}^\text{sep}_+({\phi}) \rangle & = & 2\alpha\sqrt{\eta}[\sqrt{G} \text{ sin}(\phi_1)+g\sqrt{G-1}\text{ sin}(\phi_2)] \nonumber\\
& \approx &2\alpha\sqrt{\eta}(\sqrt{G} \phi_1+g\sqrt{G-1}\phi_2) \nonumber\\
& = &2\alpha\sqrt{\eta}(\sqrt{G} +g\sqrt{G-1})\phi,
\label{eq:expsep}\\
\langle \Delta^2\hat{X}^\text{sep}_+({\phi}) \rangle & = & 2(g^2+1)(1-2\eta+2\eta G)-4g\eta\sqrt{G(G-1)}[\text{cos}(\phi_1)+\text{cos}(\phi_2)] \nonumber\\
& \approx & 2(g^2+1)(1-2\eta+2\eta G)-8g\eta\sqrt{G(G-1)}, 
\label{eq:varsep}
\end{eqnarray}
where $\phi=\beta_1\phi_1+\beta_2\phi_2$ and  $\beta_1$ and $\beta_2$ are kept the same as those used in the distributed sensing configurations.  
With Eqs.~(\ref{eq:expsep}) and~(\ref{eq:varsep}), one can show that the LOD for the separable sensing configuration with  two tSU(1,1) interferometers is given by
\begin{eqnarray}
\Delta^2\phi_\text{tSU}^\text{sep}=2 \Delta^2\phi_\text{tSU}^\text{dis}=\frac{(g^2+1)(1-2\eta+2\eta G)-4g\eta\sqrt{G(G-1)}}{2|\alpha|^2\eta(\sqrt{G}+g\sqrt{G-1})^2}.
\label{eq:LODtSUsep}
\end{eqnarray}

Next, we consider the phase sensitivity for the classical separable sensing configuration shown in Fig.~\ref{fig:s2}(b). For this, we take coherent states with the same optical power as the probe and conjugate beams used for the two tSU(1,1) interferometers, that is $\langle\hat{a}_c^\dagger\hat{a}_c\rangle=\langle \hat{a}'^\dagger_c\hat{a}'_c\rangle=G|\alpha|^2$ and $\langle \hat{b}_c^\dagger\hat{b}_c\rangle=\langle \hat{b}'^\dagger_c\hat{b}'_c\rangle=(G-1)|\alpha|^2$ with $\alpha^2\gg1$. After passing through the phase shifters and considering the losses, one can calculate the LOD following the same derivation. The LOD for the classical separable sensing configuration can then be shown to be given by,
\begin{equation}
\Delta^2\phi_\text{cla}^\text{sep}=2 \Delta^2\phi_\text{cla}^\text{dis}=\frac{(g^2+1)}{2|\alpha|^2\eta(\sqrt{G}+g\sqrt{G-1})^2}.
\label{eq:LODclasep}
\end{equation}
As can be seen, the distributed sensing configuration offers a 3 dB signal-to-noise ratio (SNR) improvement over the separable one.  It is important to note, however, that this improvement is a classical one that results from the use additional reference beams for the considered separable sensing configurations.

\section*{Phase sensitivity for distributed multi-phase sensing}
We generalize the theory for sensing $M$ distributed phases. Here, we only consider the estimation of the average of the phases, i.e. $\phi=\sum_{j=1}^{M}\phi_j/M$, with an assumption that all the phase shifts are small and the same.

\subsection*{Classical sensing scheme}
 For the classical scheme, we consider the case in which an input coherent state with optical power of $|\gamma|^2$ is divided with  a beamsplitter network (BSN) into $M$ modes of the same optical power that are used to probe $M$ unknown phase elements. After the BSN, the average photon number, $n$, hitting each of the the phase elements is given by $n=|\gamma|^2/M$. It has been shown that when $n$ photons are used to sense a phase shift, the sensitivity is given by $\Delta^2\phi_j=1/(4n)$ when there is a LO with an optical power much larger than $n$ that acts as an external phase reference~\cite{jarzyna2012quantum}. Using error propagation, one can readily show that the LOD for $\phi=\sum_{j=1}^{M}\phi_j/M$ is given by  $\Delta^2\phi_\text{classic}=1/(4Mn)$.

\subsection*{Separable scheme composed of multiple single-mode squeezed states}
An optimal quantum-enhanced separable scheme is obtained when $M$ identical and independent single-mode squeezed states are used to probe the $M$ phase elements. We consider the case in which each single mode squeezed state has an average photon number of $n$, such that the resources used to sense each phase are kept constant. In this case, the optimized sensitivity for the separable scheme has been calculated in~\cite{guo2020distributed} and is given by
\begin{equation}
\Delta^2\phi_\text{separable}=\frac{1}{4Mn(n+1)}.
\label{eq:sep}
\end{equation}
As can be seen, in this case the sensitivity shows a Heisenberg scaling in the number of photons $n$, but not in the number of modes $M$.

\subsection*{Multi-mode entangled scheme in a tSU(1,1) interferometer}
\begin{figure}[htbp]
\centering\includegraphics[width=9cm]{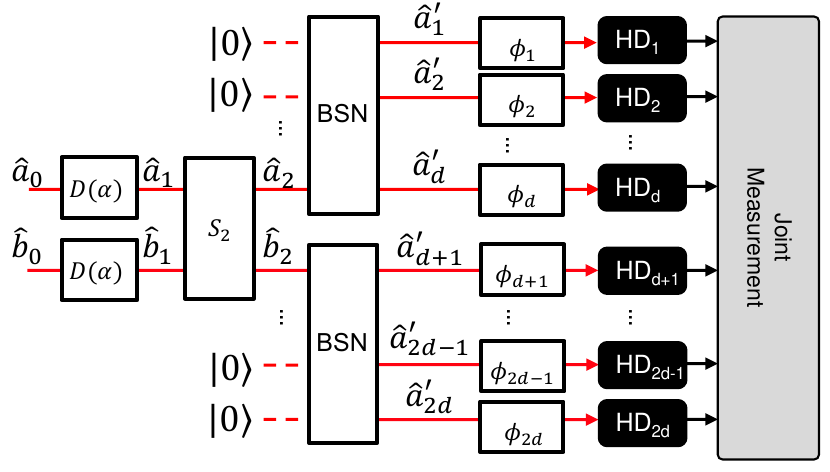}
\caption{\label{fig:6} Generalization to distributed multi-phase sensing with a combination of a tSU(1,1) interferometer and a beamsplitter network. S2: two-mode squeezing operator; BSN: beam splitter network.}
\end{figure}
To generalize the tSU(1,1) used to experimentally sense the average of two phases, we consider a configuration in which the two entangled output modes of a parametric amplifier are split into $M$ entangled modes with a BSN, as shown in Fig.~\ref{fig:6}. Here, two input modes, $\hat{a}_0$ and $\hat{b}_0$, in vacuum states, i.e. $\langle \hat{a}^\dagger_0\hat{a}_0\rangle=\langle \hat{b}^\dagger_0\hat{b}_0\rangle=0$, are displaced to become $\hat{a}_1$ and $\hat{b}_1$, respectively, with $\langle \hat{a}^\dagger_1\hat{a}_1\rangle=\langle \hat{b}^\dagger_1\hat{b}_1\rangle=|\alpha|^2$. These displaced modes then serve as the seeds for a parametric amplifier (two-mode squeezer). Note that as opposed to the experimental implementation in the two phase case, for this generalized scheme we consider seeding both input modes and assume all the involved fields have the same phase. The two output modes of the parametric amplifier, $\hat{a}_2$ and $\hat{b}_2$, follow the same nonlinear gain transformations as the one given in Eq.~(\ref{eq:amplifying}), such that the total number of photons is given by $N_{tot}=\langle \hat{a}^\dagger_2\hat{a}_2\rangle+\langle \hat{b}^\dagger_2\hat{b}_2\rangle=-2+2G+4G\alpha^2+2\alpha^2(-1+2\sqrt{G(G-1)})$. After the parametric amplifier, each mode is split into $d=M/2$ modes with a BSN. Each mode is then sent to a single phase element and experiences a phase shift $\phi_{j}$. For each mode, one can measure the generalized quadrature operator $\hat{X'}_j(\phi_j,\theta)=e^{-i\theta}\hat{a'}^\dagger_j+e^{i\theta}\hat{a'}_j$.  As done in the main manuscript, we only consider the case in which $\theta$ is set to $\pi/2$. Then, the total joint quadrature operator is defined as $\hat{X}_{+,M}(\phi)=\sum_{j=1}^M\hat{X'}_j(\phi_j)$. The expectation value and variance of this generalized distributed sensor network in the small-angle approximation can be shown to be given by:
\begin{eqnarray}
\langle \hat{X}_{+,M}({\phi}) \rangle &\approx& \frac{1}{\sqrt{M}}2\sqrt{2}\alpha(\sqrt{G}+\sqrt{G-1})\sum_{j=1}^M\phi_j
=2\sqrt{2}\sqrt{M}\alpha(\sqrt{G}+\sqrt{G-1})\phi,
\label{eq:genexp}\\
\langle\Delta^2\hat{X}_{+,M}({\phi})\rangle&\approx& M(-1+2G-2\sqrt{G(G-1)}).
\label{eq:vargen}
\end{eqnarray}
Equations~(\ref{eq:genexp}) and (\ref{eq:vargen}) can then be used to obtain the LOD for the multi-phase sensing scheme based on a multi-mode entangled stated obtained with a tSU(1,1) interferometer and a BSN, such that
\begin{eqnarray}
\Delta^2\phi_\text{ent}=\frac{-1+2G-2\sqrt{G(G-1)}}{8|\alpha|^2(\sqrt{G}+\sqrt{G-1})^2}.
\label{eq:genLOD}
\end{eqnarray}
Note that Eq.~(\ref{eq:genLOD}) does not explicitly depend on the number of modes $M$. To better compare with the other schemes for multi-phase estimation, we use the constraint that the average number of photons hitting each phase element is $n$, i.e. $N_{tot}/M=\langle \hat{a}'^\dagger_j\hat{a}'_j\rangle=n$. To optimize the sensitivity for a fixed photon number incident on each phase shifter, we follow the procedure in~\cite{guo2020distributed}  based on Lagrangian multipliers with this constraint. We can construct the Lagrange function as
\begin{equation}
\mathcal{L}(\alpha,G,\lambda)=\Delta\phi_\text{ent}+\lambda [(N_{tot}/M)-n].
\label{eq:lag}
\end{equation}
By substituting the solution of Eq.~(\ref{eq:lag}) into the Eq.~(\ref{eq:genLOD}), the optimal LOD becomes 
\begin{equation}
\Delta^2\phi_\text{ent}=\frac{1}{2Mn(Mn+2)}.
\label{eq:optLOD}
\end{equation}
As can be seen, for this scheme in which entanglement is distributed to all phase elements, Heisenberg scaling for both the number of photons $n$ and the number of modes $M$ is achieved.

\section*{Experimental details}

\begin{figure}[htbp]
\centering\includegraphics[width=13cm]{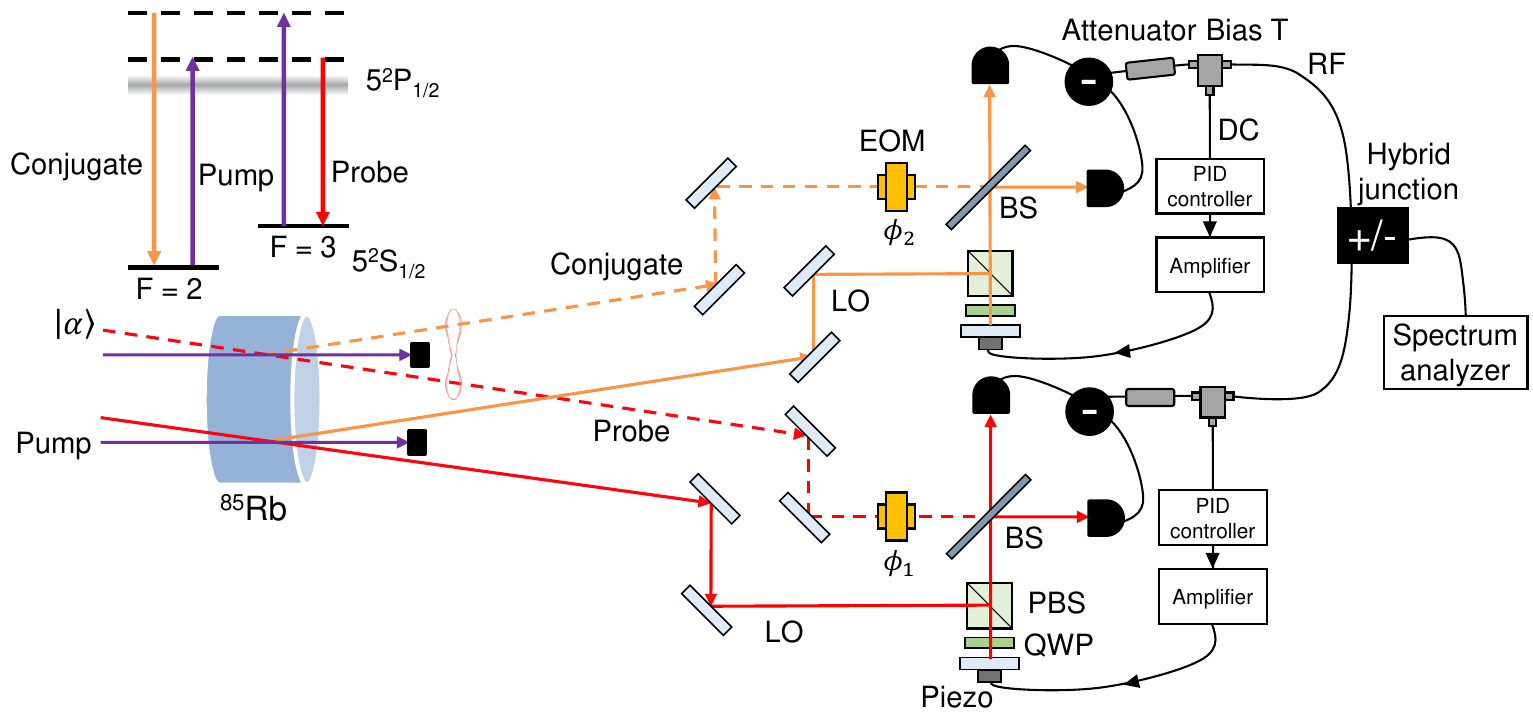}
\caption{\label{fig:s5} Experimental setup for the  implementation of  the distributed two-phase sensing configuration with a tSU(1,1) interferometer. LO: local oscillator; EOM: electro-optic modulator; PBS: polarizing beamsplitter; QWP: quarter waveplate; BS: beamsplitter.}
\end{figure}

We construct the tSU(1,1) interferometer using a FWM process in a double-lambda configuration in a $^{85}$Rb cell, as shown in Fig.~\ref{fig:s5}~\cite{pooser2020truncated}. The FWM process is a coherent process in which two photons from a strong pump are absorbed to generate probe and conjugate photons. The frequency of the probe is redshifted by 3.044~GHz with respect to the pump frequency by double passing an acousto-optic modulator (AOM) set to impart a frequency shift of 1.522~GHz. The pump and probe beams are overlapped at an angle of $0.34^\circ$ in a $^{85}$Rb cell with 1/$e^2$ beam diameters of $920~\mu$m and $560~\mu$m, respectively. In our configuration, a pump with 360~mW of optical power leads to a FWM with a gain of $G\approx 5$. A spatially separated FWM process in the same vapor cell is used to generate LOs that have the same spatial modes as the probe and conjugate beams for use in the dual balanced homodyne detection.

For the dual homodyne detection, the probe and conjugate are interfered with their corresponding LOs using a 50:50 beamsplitter and are then measured by a pair of balanced photodetectors (PDs). The PD difference signals for the probe and conjugate are sent to variable attenuators to implement the classical gain/attenuation $g$. Note that we moderately adjusted $g$ to make $\beta_1=\beta_2=1/2$ in the experiment. In order to set the LO phases such that $\theta_1=\theta_2=\pi/2$, a bias-tee is used to divide the electric signal from each of the PDs into a low frequency (DC) and a high frequency (RF) component with a cutoff frequency of 20 kHz, with the DC component used to lock the difference signal at the zero-crossing with a piezo-driven mirror. The RF signals, which contain the squeezing and entanglement information, are sent to a hybrid junction to obtain the joint quadrature operators. The hybrid junction outputs both the sum and  difference signals of the probe and conjugate homodyne detectors, with the phase sum signal used here. 

In order to introduce phase modulations that serve as the signals, two electro-optic modulators (EOMs) are driven at a frequency of 300 kHz with independently controlled voltages to implement the different experimental configurations. For the data in Fig.~3(a) of the main text, the EOMs were driven with RF drivers with an applied peak-to-peak voltage of 30 mV, while for the data in Fig.~3(b) of the main text the peak-to-peak voltage was reduced to 10 mV to impart a smaller signal. 

\section*{SNR Data Analysis}
To obtain an accurate measure of the improvement in SNR between the separable and distributed sensing configurations, we follow the procedure outlined in~\cite{hill1990accurate}. We start by using the spectrum analyzer to directly measure the signal, which contains contributions from both the actual signal and the noise, as well as the noise floor. To estimate the actual SNR, we need to consider a correction factor (CF) that depends on the measured SNR as outlined in~\cite{hill1990accurate}. The actual SNR in dB is then given by: SNR (dB) = Measured signal (dBm) - Measured noise (dBm) - CF. This equation leads to actual SNRs of $-0.21 \pm 0.42$ dB for the separable sensing configuration and $2.80\pm0.31$ dB for the distributed sensing configuration. Thus, the SNR improvement of the distributed configuration over the separable one is $3.0\pm0.5$~dB, as expected.


\begin{thebibliography}{10}
\newcommand{\enquote}[1]{``#1''}

\bibitem{caves1981quantum}
C.~M. Caves, \enquote{Quantum-mechanical noise in an interferometer,}
  {\protect\JournalTitle{Physical Review D}} \textbf{23}, 1693 (1981).

\bibitem{giovannetti2004quantum}
V.~Giovannetti, S.~Lloyd, and L.~Maccone, \enquote{Quantum-enhanced
  measurements: beating the standard quantum limit,}
  {\protect\JournalTitle{Science}} \textbf{306}, 1330--1336 (2004).

\bibitem{pirandola2018advances}
S.~Pirandola, B.~R. Bardhan, T.~Gehring, C.~Weedbrook, and S.~Lloyd,
  \enquote{Advances in photonic quantum sensing,} {\protect\JournalTitle{Nat.
  Photonics}} \textbf{12}, 724--733 (2018).

\bibitem{lawrie2019quantum}
B.~J. Lawrie, P.~D. Lett, A.~M. Marino, and R.~C. Pooser, \enquote{Quantum
  sensing with squeezed light,} {\protect\JournalTitle{ACS Photonics}}
  \textbf{6}, 1307--1318 (2019).

\bibitem{RevModPhys.90.035005}
L.~Pezz\`e, A.~Smerzi, M.~K. Oberthaler, R.~Schmied, and P.~Treutlein,
  \enquote{Quantum metrology with nonclassical states of atomic ensembles,}
  {\protect\JournalTitle{Rev. Mod. Phys.}} \textbf{90}, 035005 (2018).

\bibitem{RevModPhys.89.035002}
C.~L. Degen, F.~Reinhard, and P.~Cappellaro, \enquote{Quantum sensing,}
  {\protect\JournalTitle{Rev. Mod. Phys.}} \textbf{89}, 035002 (2017).

\bibitem{anderson2017phase}
B.~E. Anderson, P.~Gupta, B.~L. Schmittberger, T.~Horrom, C.~Hermann-Avigliano,
  K.~M. Jones, and P.~D. Lett, \enquote{Phase sensing beyond the standard
  quantum limit with a variation on the {SU(1,1)} interferometer,}
  {\protect\JournalTitle{Optica}} \textbf{4}, 752--756 (2017).

\bibitem{gupta2018optimized}
P.~Gupta, B.~L. Schmittberger, B.~E. Anderson, K.~M. Jones, and P.~D. Lett,
  \enquote{Optimized phase sensing in a truncated {SU(1,1)} interferometer,}
  {\protect\JournalTitle{Opt. Express}} \textbf{26}, 391--401 (2018).

\bibitem{prajapati2019polarization}
N.~Prajapati and I.~Novikova, \enquote{Polarization-based truncated {SU(1,1)}
  interferometer based on four-wave mixing in {Rb} vapor,}
  {\protect\JournalTitle{Opt. Lett.}} \textbf{44}, 5921--5924 (2019).

\bibitem{pooser_ultrasensitive_2015}
R.~C. Pooser and B.~Lawrie, \enquote{Ultrasensitive measurement of
  microcantilever displacement below the shot-noise limit,}
  {\protect\JournalTitle{Optica}} \textbf{2}, 393--399 (2015).

\bibitem{pooser2020truncated}
R.~Pooser, N.~Savino, E.~Batson, J.~Beckey, J.~Garcia, and B.~Lawrie,
  \enquote{Truncated nonlinear interferometry for quantum-enhanced atomic force
  microscopy,} {\protect\JournalTitle{Phys. Rev. Lett.}} \textbf{124}, 230504
  (2020).

\bibitem{xia2023entanglement}
Y.~Xia, A.~R. Agrawal, C.~M. Pluchar, A.~J. Brady, Z.~Liu, Q.~Zhuang, D.~J.
  Wilson, and Z.~Zhang, \enquote{Entanglement-enhanced optomechanical sensing,}
  {\protect\JournalTitle{Nat. Photonics}} \textbf{17}, 470--477 (2023).

\bibitem{caves_reframing_2020}
C.~M. Caves, \enquote{Reframing {SU}(1,1) interferometry,}
  {\protect\JournalTitle{Adv. Quantum Technol.}} p. 1900138 (2020).

\bibitem{moreau2019imaging}
P.-A. Moreau, E.~Toninelli, T.~Gregory, and M.~J. Padgett, \enquote{Imaging
  with quantum states of light,} {\protect\JournalTitle{Nat. Rev. Phys.}}
  \textbf{1}, 367--380 (2019).

\bibitem{ono2013entanglement}
T.~Ono, R.~Okamoto, and S.~Takeuchi, \enquote{An entanglement-enhanced
  microscope,} {\protect\JournalTitle{Nat. Commun.}} \textbf{4}, 2426 (2013).

\bibitem{hong2021quantum}
S.~Hong, Y.-S. Kim, Y.-W. Cho, S.-W. Lee, H.~Jung, S.~Moon, S.-W. Han, H.-T.
  Lim \emph{et~al.}, \enquote{Quantum enhanced multiple-phase estimation with
  multi-mode {N00N} states,} {\protect\JournalTitle{Nat. Commun.}} \textbf{12},
  5211 (2021).

\bibitem{hong2022practical}
S.~Hong, J.~u. Rehman, Y.-S. Kim, Y.-W. Cho, S.-W. Lee, S.-Y. Lee, and H.-T.
  Lim, \enquote{Practical sensitivity bound for multiple phase estimation with
  multi-mode {N00N} states,} {\protect\JournalTitle{Laser Photonics Rev.}}
  \textbf{16}, 2100682 (2022).

\bibitem{zhuang2018distributed}
Q.~Zhuang, Z.~Zhang, and J.~H. Shapiro, \enquote{Distributed quantum sensing
  using continuous-variable multipartite entanglement,}
  {\protect\JournalTitle{Phys. Rev. A}} \textbf{97}, 032329 (2018).

\bibitem{proctor2018multiparameter}
T.~J. Proctor, P.~A. Knott, and J.~A. Dunningham, \enquote{Multiparameter
  estimation in networked quantum sensors,} {\protect\JournalTitle{Phys. Rev.
  Lett.}} \textbf{120}, 080501 (2018).

\bibitem{guo2020distributed}
X.~Guo, C.~R. Breum, J.~Borregaard, S.~Izumi, M.~V. Larsen, T.~Gehring,
  M.~Christandl, J.~S. Neergaard-Nielsen, and U.~L. Andersen,
  \enquote{Distributed quantum sensing in a continuous-variable entangled
  network,} {\protect\JournalTitle{Nat. Phys.}} \textbf{16}, 281--284 (2020).

\bibitem{zhao2021field}
S.-R. Zhao, Y.-Z. Zhang, W.-Z. Liu, J.-Y. Guan, W.~Zhang, C.-L. Li, B.~Bai,
  M.-H. Li, Y.~Liu, L.~You \emph{et~al.}, \enquote{Field demonstration of
  distributed quantum sensing without post-selection,}
  {\protect\JournalTitle{Phys. Rev. X}} \textbf{11}, 031009 (2021).

\bibitem{kim2023distributed}
D.-H. Kim, S.~Hong, Y.-S. Kim, Y.~Kim, R.~Lee, Seung-Woo~Pooser, K.~Oh, S.-Y.
  Lee, C.~Lee, and H.-T. Lim, \enquote{Distributed quantum sensing of multiple
  phases with fewer photons,} {\protect\JournalTitle{Nat. Commun.}}
  \textbf{15}, 266 (2024).

\bibitem{qi2018ultimate}
H.~Qi, K.~Br{\'a}dler, C.~Weedbrook, and S.~Guha, \enquote{Quantum precision of
  beam pointing,} {\protect\JournalTitle{arXiv:1808.01302}}  (2018).

\bibitem{komar2014quantum}
P.~Komar, E.~M. Kessler, M.~Bishof, L.~Jiang, A.~S. S{\o}rensen, J.~Ye, and
  M.~D. Lukin, \enquote{A quantum network of clocks,}
  {\protect\JournalTitle{Nat. Phys.}} \textbf{10}, 582--587 (2014).

\bibitem{derevianko2018detecting}
A.~Derevianko, \enquote{Detecting dark-matter waves with a network of
  precision-measurement tools,} {\protect\JournalTitle{Phys. Rev. A}}
  \textbf{97}, 042506 (2018).

\bibitem{carney2020proposal}
D.~Carney, S.~Ghosh, G.~Krnjaic, and J.~M. Taylor, \enquote{Proposal for
  gravitational direct detection of dark matter,} {\protect\JournalTitle{Phys.
  Rev. D}} \textbf{102}, 072003 (2020).

\bibitem{carney2021mechanical}
D.~Carney, G.~Krnjaic, D.~C. Moore, C.~A. Regal, G.~Afek, S.~Bhave,
  B.~Brubaker, T.~Corbitt, J.~Cripe, N.~Crisosto \emph{et~al.},
  \enquote{Mechanical quantum sensing in the search for dark matter,}
  {\protect\JournalTitle{Quantum Sci. Technol.}} \textbf{6}, 024002 (2021).

\bibitem{attanasio2022snowmass}
A.~Attanasio, S.~A. Bhave, C.~Blanco, D.~Carney, M.~Demarteau, B.~Elshimy,
  M.~Febbraro, M.~A. Feldman, S.~Ghosh, A.~Hickin \emph{et~al.},
  \enquote{Snowmass 2021 white paper: The windchime project,}
  {\protect\JournalTitle{arXiv:2203.07242}}  (2022).

\bibitem{hudelist2014quantum}
F.~Hudelist, J.~Kong, C.~Liu, J.~Jing, Z.~Ou, and W.~Zhang, \enquote{Quantum
  metrology with parametric amplifier-based photon correlation
  interferometers,} {\protect\JournalTitle{Nat. Commun.}} \textbf{5}, 1--6
  (2014).

\bibitem{marino2009tunable}
A.~M. Marino, R.~C. Pooser, V.~Boyer, and P.~D. Lett, \enquote{Tunable delay of
  {E}instein--{P}odolsky--{R}osen entanglement,}
  {\protect\JournalTitle{Nature}} \textbf{457}, 859--862 (2009).

\bibitem{anderson2017optimal}
B.~E. Anderson, B.~L. Schmittberger, P.~Gupta, K.~M. Jones, and P.~D. Lett,
  \enquote{Optimal phase measurements with bright-and vacuum-seeded {SU(1,1)}
  interferometers,} {\protect\JournalTitle{Phys. Rev. A}} \textbf{95}, 063843
  (2017).

\bibitem{liu2018quantum}
S.~Liu, Y.~Lou, J.~Xin, and J.~Jing, \enquote{Quantum enhancement of phase
  sensitivity for the bright-seeded {SU(1,1)} interferometer with direct
  intensity detection,} {\protect\JournalTitle{Phys. Rev. Appl.}} \textbf{10},
  064046 (2018).

\bibitem{ge2018distributed}
W.~Ge, K.~Jacobs, Z.~Eldredge, A.~V. Gorshkov, and M.~Foss-Feig,
  \enquote{Distributed quantum metrology with linear networks and separable
  inputs,} {\protect\JournalTitle{Phys. Rev. Lett.}} \textbf{121}, 043604
  (2018).

\bibitem{oh2022distributed}
C.~Oh, L.~Jiang, and C.~Lee, \enquote{Distributed quantum phase sensing for
  arbitrary positive and negative weights,} {\protect\JournalTitle{Phys. Rev.
  Res.}} \textbf{4}, 023164 (2022).

\bibitem{pai2022magneto}
Y.-Y. Pai, C.~E. Marvinney, C.~Hua, R.~C. Pooser, and B.~J. Lawrie,
  \enquote{Magneto-optical sensing beyond the shot noise limit,}
  {\protect\JournalTitle{Adv. Quantum Technol.}} \textbf{5}, 2100107 (2022).

\bibitem{mccormick_strong_2007}
C.~F. McCormick, V.~Boyer, E.~Arimondo, and P.~D. Lett, \enquote{Strong
  relative intensity squeezing by four-wave mixing in rubidium vapor,}
  {\protect\JournalTitle{Opt. Lett.}} \textbf{32}, 178--180 (2007).

\bibitem{PhysRevA.78.043816}
C.~F. McCormick, A.~M. Marino, V.~Boyer, and P.~D. Lett, \enquote{Strong
  low-frequency quantum correlations from a four-wave-mixing amplifier,}
  {\protect\JournalTitle{Phys. Rev. A}} \textbf{78}, 043816 (2008).

\bibitem{woodworth2020transmission}
T.~S. Woodworth, K.~W.~C. Chan, C.~Hermann-Avigliano, and A.~M. Marino,
  \enquote{Transmission estimation at the {C}ram{\'e}r-{R}ao bound for squeezed
  states of light in the presence of loss and imperfect detection,}
  {\protect\JournalTitle{Phys. Rev. A}} \textbf{102}, 052603 (2020).

\bibitem{woodworth2022transmission}
T.~S. Woodworth, C.~Hermann-Avigliano, K.~W.~C. Chan, and A.~M. Marino,
  \enquote{Transmission estimation at the fundamental quantum
  {C}ram{\'e}r-{R}ao bound with macroscopic quantum light,}
  {\protect\JournalTitle{EPJ Quantum Technol.}} \textbf{9}, 38 (2022).

\bibitem{you2019conclusive}
C.~You, S.~Adhikari, X.~Ma, M.~Sasaki, M.~Takeoka, and J.~P. Dowling,
  \enquote{Conclusive precision bounds for {SU(1,1)} interferometers,}
  {\protect\JournalTitle{Phys. Rev. A}} \textbf{99}, 042122 (2019).

\bibitem{boyer2008}
V.~Boyer, A.~M. Marino, R.~C. Pooser, and P.~D. Lett, \enquote{Entangled images
  from four-wave mixing,} {\protect\JournalTitle{Science}} \textbf{321},
  544--547 (2008).

\bibitem{pooser2014continuous}
R.~Pooser and J.~Jing, \enquote{Continuous-variable cluster-state generation
  over the optical spatial mode comb,} {\protect\JournalTitle{Phys. Rev. A}}
  \textbf{90}, 043841 (2014).
  
\bibitem{jasperse2011relative}
M.~Jasperse, L.~Turner, and R.~Scholten, \enquote{Relative intensity squeezing
  by four-wave mixing with loss: an analytic model and experimental
  diagnostic,} {\protect\JournalTitle{Optics Express}} \textbf{19}, 3765--3774
  (2011).
  
\bibitem{vsafranek2018estimation}
D.~{\v{S}}afr{\'a}nek, \enquote{Estimation of gaussian quantum states,}
  {\protect\JournalTitle{Journal of Physics A: Mathematical and Theoretical}}
  \textbf{52}, 035304 (2018).
  
\bibitem{jarzyna2012quantum}
M.~Jarzyna and R.~Demkowicz-Dobrza{\'n}ski, \enquote{Quantum interferometry
  with and without an external phase reference,}
  {\protect\JournalTitle{Physical Review A}} \textbf{85}, 011801 (2012).
  
\bibitem{hill1990accurate}
D.~Hill and D.~Haworth, \enquote{Accurate measurement of low signal-to-noise
  ratios using a superheterodyne spectrum analyzer,}
  {\protect\JournalTitle{IEEE Trans. Instrum. Meas.}} \textbf{39}, 432--435
  (1990).

\end{thebibliography}
\end{document}